\documentclass[aps,preprint,nofootinbib]{revtex4}
\usepackage{graphicx}
\usepackage{epstopdf}
\usepackage{amsmath}
\usepackage{amssymb}
\usepackage[greek,english]{babel}
\usepackage[iso-8859-7]{inputenc}
\usepackage{color}

\def\ba{\begin{eqnarray}}
\def\ea{\end{eqnarray}}
\def\be{\begin{equation}}
\def\ee{\end{equation}}

\def\gtorder{\mathrel{\raise.3ex\hbox{$>$}\mkern-14mu
             \lower0.6ex\hbox{$\sim$}}}
\def\ltorder{\mathrel{\raise.3ex\hbox{$<$}\mkern-14mu
             \lower0.6ex\hbox{$\sim$}}}

\def\dalemb#1#2{{\vbox{\hrule height.#2pt
  \hbox{\vrule width.#2pt height#1pt \kern#1pt \vrule width.#2pt}
    \hrule height.#2pt}}}

\begin{document}


\rightline{T$\Theta\Delta$}

\title{A Bayesian approach to the probability of coronary heart disease 
subject to 
the --308 tumor necrosis factor-$\alpha$ SNP
}

\author{Ekaterini~Vourvouhaki${}^{1}$\footnote{Email: kvourvouhaki@gmail.com} 
and C.~Sofia~Carvalho${}^{2,3}$\footnote{Email: carvalho.c@gmail.com} }
\affiliation{${}^{1}$ 
Department of Science in Dietetics and Nutrition, 
Harokopio University of Athens, 
Eleutheriou Venizelou 70, 
GR 176~71 Athens, Greece}
\affiliation{${}^{2}$ 
Instituto de Plasmas e Fus\~ao Nuclear, Instituto Superior T\'ecnico,
Av. Rovisco Pais, 1, 1049-001 Lisboa, Portugal}
\affiliation{${}^{3}$
Academy of Athens, Research Center for Astronomy and Applied Mathematics, Soranou Efessiou 4, 11-527, Athens, Greece
}
\begin{abstract}
We study the correlation of the occurrence of coronary heart disease (CHD) with the presence of the single-nucleotide polymorphism (SNP) at the -308 position of the tumor necrosis factor alpha (TNF-$\alpha$) gene. We also consider the influence of the occurrence of type 2 diabetes (t2DM).
Using Bayesian inference, we first pursue a bottom-up approach to compute the working hypothesis and the probabilities derivable from the data. We then pursue a top-down approach by modelling the signal pathway that causally connects the SNP with the emergence of CHD. We compute the functional form of the probability of CHD conditional on the presence of the SNP in terms of both the statistical and biochemical properties of the system. 
From the  probability of occurrence of a disease conditional on a given risk factor, we explore the possibility of extracting information on the pathways involved in the occurrence of the disease. 
This is a first study that we want to systematise into a comprehensive formalism to be applied to the inference of the mechanism connecting the risk factors to the disease.
\end{abstract}

\date{\today}


\maketitle

\section{Introduction}

We are interested in the association of diseases, in particular of
coronary heart disease (CHD), with genetic factors in order to determine
underlying genetically-driven 
functional mechanisms that are causally related to the
disease. In this context, environmental factors are regarded as contaminants.
Among the risk factors for the emergence of CHD, genetic determinants 
may provide a wealth of information on the nature of the disease, which can be used to develop new diagnosis and treatment methods. The study of these factors has attracted the effort of many research teams for the identification of disease susceptibility genes as well as acquired somatic mutations. 
Among these genes is that of the tumor necrosis factor alpha (TNF-$\alpha$), a pleiotropic cytokine produced mainly by macrophages and T-cells which is involved in the inflammatory response of the immune system \cite{vassali_1992}.

It has been suggested that the TNF-$\alpha$
gene affects the modulation of lipid metabolism, obesity
susceptibility, and insulin resistance \cite{vendrell_2003,dedoussis_2005,elahi_2008}, thus being potentially implicated in the
development of cardiovascular diseases 
(see Ref.~\cite{vourvouhaki_2008} and references therein). 
Several single-nucleotide polymorphisms (SNPs) have been identified in the human TNF-$\alpha$ \cite{westendorp_1997,abraham_1999}.
The best documented of these SNPs are at position -308 of the TNF-$\alpha$ gene promoter. It involves the substitution of guanine (G) for adenine (A) and the creation of two alleles, TNF1(A) and TNF2(G), and three genotypes, GG, GA and AA \cite{wilson_1992}.
There is evidence implicating TNF-$\alpha$ in an increased susceptibility to the pathogenesis of a variety of diseases (see Ref.~\cite{elahi_2008} and references therein).
However, the results on its association with CHD are contradictory, some implying different influence of the two alleles on the prevalence of CHD \cite{dedoussis_2005,elahi_2008}, others implying no association \cite{kammia_sxesh_1,kammia_sxesh_2,kammia_sxesh_3,kammia_sxesh_4}. This conflict is due in part to the results being based on the frequentist analysis \cite{stephens_2009}.

In order to infer the risk of CHD derived from potential risk factors, it is important to develop a formalism that extracts all possible information from the data and combines them with other data sets on different intervening factors for a consistent inference of the correlations. Here we introduce a possible formalism based on Bayesian inference and test its applicability on the three-variable data set from Ref.~\cite{vendrell_2003}. 
In this manuscript we attempt to quantify the risk of occurrence of CHD based on its association with the SNP at the position -308 of the TNF-$\alpha$ promoter. This entails the calculation of a probability distribution for the occurrence of CHD conditional on the SNP or other factors. 
The influence of other factors is here illustrated by the occurrence of type 2 diabetes (t2DM).

When, instead of computing the probability distribution of some
quantity produced by the process, we compute the conditional 
probability of an unsolved variable in the process given the observed
variables, we are solving an inverse probability problem. This
requires the use of the Bayes theorem.
The Bayesian approach has been used extensively for parameter inference and model selection from cosmology \cite{brewer_bay,trotta_2005,bridges_2008} to biology \cite{imoto_2004,girolami_2008,stumpf_2009,stumpf_2009b} 
among many others.
In this case, we observe the occurrence of a given disease and the
correlation with a SNP.
In the absence of a theory, we 
relate the SNP with the
disease via a model of the potentially implicated pathways. The
parameters of this model are the relevant factors that we want to
infer from the data. 
Such a theory would be important as a first step to predict the occurrence of a genetically-driven disease for a given polymorphism,
as well as to understand the mechanism
of genetic mutation from which polymorphisms derive.
In this study we propose a solution to the first problem and will approach the second problem in a forthcoming study.

The causal relation between the risk factors and the occurrence of the disease is a function of the rates which characterize the implicated pathways.
Here, knowing how the occurrence of the disease is distributed over the parameter space of the risk factors and knowing how the risk factors act at the biochemical level, we show how we can extract information on the pathway involved in the emergence of the disease. 
However, since the pathways involved are most likely interconnected with others, sourced by different factors, the next step would be to allow for the participation of various factors 
in the emergence of the same disease.
There is a plethora of sparse phenomenological/symptomatic data on the simultaneous occurrence of SNP's and diseases from which correlations are tentatively drawn.  
The formalism developed here can be extended to other factors in the effort to systematise the sparse data to identify risk factors, to combine them into a comprehensive model for the mechanism that leads to the disease and therefrom to infer a universal law of gene mutation. 




The manuscript is organized as follows. 
In section \ref{section:bottom-up} 
we select the working hypothesis for the relation between the SNP and CHD on the basis of the Bayes factors and compute the probabilities in the presence of the SNP derivable from the data. One of these will be used as the likelihood for the occurrence of CHD. In section \ref{section:top-down} we suggest a simplistic model for the signalling pathway between the onset of the SNP and the emergence of CHD, and compute the posterior probability for the occurrence of CHD. Finally we comment on the results and indicate the research routes that we will be exploring next.

\section{Bottom-up approach}\label{section:bottom-up}


We will base our analysis on the data reported in Ref.~\cite{vendrell_2003}, which consist of frequencies of occurrence of CHD as a function of the SNP at the position $-308$ of the TNF-$\alpha$ promoter. It is also advanced a correlation between the SNP and an increased predisposition to CHD in type 2 diabetic patients. 
That study also comprises an analysis of the gender dependence on diabetes. 

The sample of CHD patients consists of $N_{CHD}=341$ randomly selected patients. Out of these, $N_{CHD,t2DM}=106$ also suffered from t2DM. Another sample of type 2 diabetic patients numbering 
$N_{\overline{CHD},t2DM}=135$ was
selected among non-CHD patients. These two samples, together with a
control sample of $N_{\overline{CHD},\overline{t2DM}}=207$ 
non-CHD non-diabetic patients, were analysed for the
occurrence of the $-308$ TNF-$\alpha$ SNP. Thus the total number of
diabetic patients consists of a random and a non-random
component on the factor CHD. 
Since we are interested in studying the
correlation between the SNP and CHD, we cannot use the data on the
sample of non-CHD diabetic patients to extract information on the frequency of
occurrence of CHD given that diabetes had occurred, because the information would be biased. 
We can, however, derive information on the frequency of diabetes given
that the SNP or CHD occurred. 
The data are summarized in Table~\ref{table:ena}.

\begin{table}[t]
\begin{tabular}{ccccc}
&\multicolumn{2}{c}{$CHD$}& \multicolumn{2}{c}{$\overline{CHD}$}\\ \cline{2-5}
~&~ $t2DM$ ~&~ $\overline{t2DM}$ ~&~ $t2DM$ ~&~ $\overline{t2DM}$\\ \hline
$SNP$ &43 &67 &26 & 48 \\ 
$\overline{SNP}$ &63 &168 &109 &159\\ \hline
\end{tabular}
\caption{\label{table:ena} {\bf The data.} Frequencies of the TNF$\alpha$--308 SNP in CHD patients, t2DM patients and controls.}
\end{table}

\subsection{Model comparison: the Bayesian evidence}

\begin{figure}[t]
\setlength{\unitlength}{1.cm}
\includegraphics[width=12.5cm]{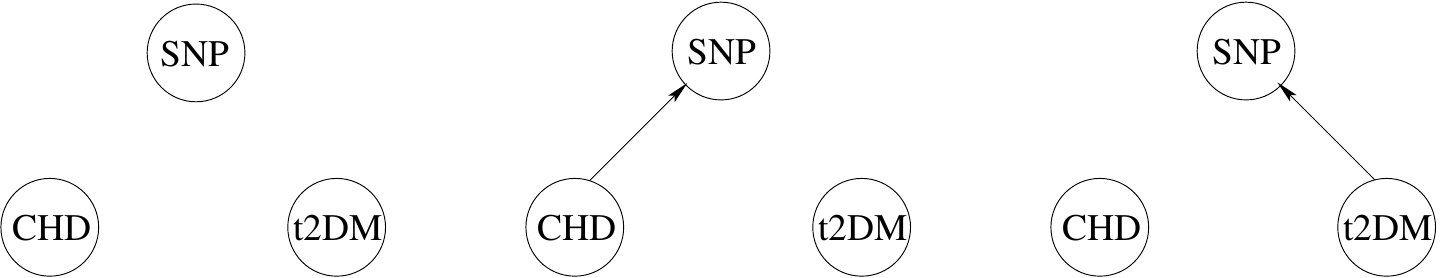}
\put(-12.5,2.5){\framebox[1cm]{$H_{00}$}}
\put(-7.9,2.5){\framebox[1cm]{$H_{01}$}}
\put(-3.4,2.5){\framebox[1cm]{$H_{10}$}}
\vfill
\vspace{1.cm}
\includegraphics[width=12.5cm]{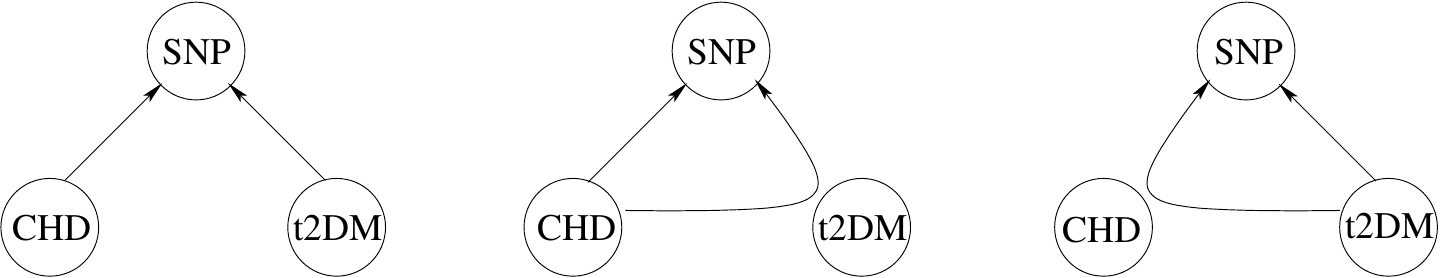}
\put(-12.5,2.5){\framebox[1cm]{$H_{11}$}}
\put(-7.9,2.5){\framebox[1cm]{$H_{11}^{chd}$}}
\put(-3.4,2.5){\framebox[1cm]{$H_{11}^{t2dm}$}}
\caption{\small \baselineskip=0.5cm{
{\bf Diagram of the hypotheses considered. }
The circles denote the corresponding variables. The arrows denote a correlation between the encircled variables which they connect. A straight arrow indicates that the probability of the variable at the tip of the arrow is conditional on the variable at the origin of the arrow. A curved arrow indicates that the probability of the variable at the tip of the arrow is conditional on the variable at the inflection of the arrow, which moreover is conditional on the variable at the origin of the arrow.
In the upper set we have $H_{00},$ $H_{01}$ and $H_{10};$ 
in the lower set we have $H_{11},$
$H_{11}^{chd}$ and $H_{11}^{t2dm}.$}}
\label{fig:pikassoyla1}
\end{figure}
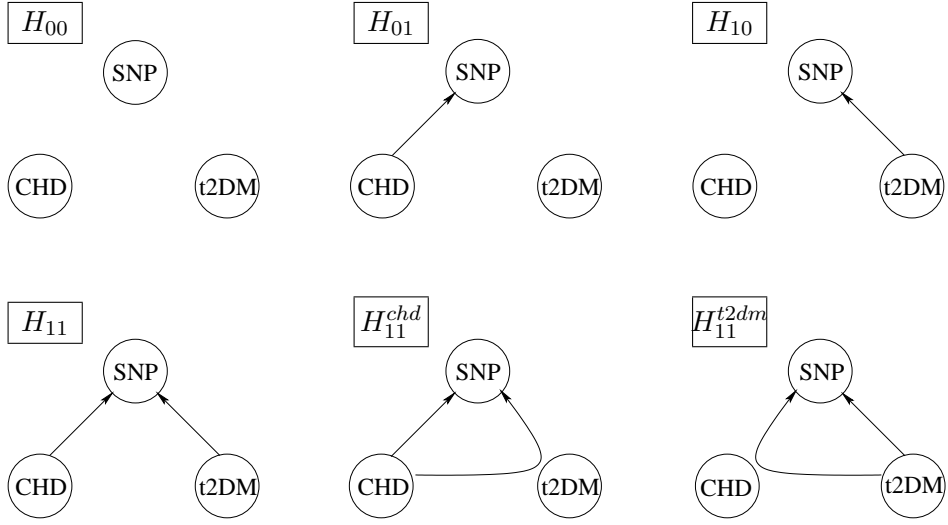

Given the data, we can derive the influence of CHD and t2DM on
the SNP. We have three variables, namely occurrence of CHD ($CHD$), occurrence
of t2DM ($t2DM$) and presence of the SNP ($SNP$), 
and six hypotheses 
for the presence of the SNP. The hypotheses are the following:
$H_{00}$: the probability of the SNP does not depend on the occurrence of
either CHD or t2DM;
$H_{01}$: the probability of the SNP depends on the occurrence of CHD;
$H_{10}$: the probability of the SNP depends on the occurrence of t2DM; 
$H_{11}$: the probability of the SNP depends on the independent
occurrence of both CHD and t2DM;
$H_{11}^{chd}$: the probability of the SNP depends on the occurrence
of CHD and on the occurrence of t2DM given that CHD
is present; 
$H_{11}^{t2dn}$: the probability of the SNP depends on the occurrence
of t2DM and on the occurrence of CHD given that diabetes is present.
These are schematically depicted in Fig.~\ref{fig:pikassoyla1}

We note that, given the selection criterion for the population of 
$N_{\overline{CHD},t2DM},$ we cannot use the corresponding data to
infer on the occurrence of CHD given the occurrence of t2DM 
since they would bias the results.
For this reason, $H_{11}^{t2dm}$ is excluded as a viable hypothesis given the data collected.
We proceed to compare the remaining hypotheses based on Bayesian evidence.
The probability of an hypothesis given the data is the
posterior probability of the corresponding model \cite{mackay}
\ba
P(H_{i}\vert D)= { {P(D\vert H_{i})P(H_{i})}\over P(D)}
\ea
where $ P(D\vert H_{i})$ is the evidence, $P(H_{i})$ is the prior probability of $H_{i}$
and $P(D)=\sum_{i}P(D\vert H_{i})P(H_{i}).$
In order to infer which hypothesis is more likely in view of the data, we
compare the evidence computed for the alternative hypotheses. The
evidence is the integral of the likelihood over the parameter space
$\theta$ of the model 
\ba
P(D\vert H_{i})
=\int d\theta P(D\vert \theta \wedge H_{i})P(\theta\vert H_{i}).
\ea
Assuming equal prior probabilities for the different hypotheses, then
\ba
{P(H_{i}\vert D)\over P(H_{j}\vert D)}
={P(D\vert H_{i})\over P(D\vert H_{j})}.
\ea
We compute the evidence for the five hypotheses described above (for
details see Appendix \ref{app:evidence}). In order to compare the hypotheses,
we take the logarithm of the ratio of the corresponding
evidences, $P(H_{i}\vert D)/P(H_{j}\vert D)=B_{ij},$ 
which we present in Table~\ref{table:dyo}.  This quantity is known as
the Bayes factor and gives empirical levels of significance for the
strength of the evidence.
It also encapsulates Occam's factor
which measures the adequacy of the hypothesis to the data over the
parameter space of the hypothesis \cite{mackay}.  
The levels of significance ascribed to the Bayes
factor are calibrated by the Jeffrey's scale \cite{jeffrey} as
follows: if $1.0<B_{ij}<3.2,$ $H_{i}$ should not be favoured over $H_{j};$
if $3.2<B_{ij}<10,$ there is substantial evidence for $H_{i}$ over
$H_{j};$ if $10<B_{ij}<100$ there is strong evidence, while for
$B_{ij}>100$ the evidence for $H_{ij}$ should be considered decisive.
In the first column, we find the Bayes factors which relate each
hypothesis with $H_{00}.$  Since hypothesis $H_{00}$ describes the data as the result of a random process, this column measures the preference for a departure from randomness \cite{frank_2009}.
From these values we infer that all hypotheses
are substantially favoured over $H_{00}.$
The values seem to suggest that $H_{11}^{chd}$ 
is also favoured over the other hypotheses, however they are not sufficient to infer substantial evidence. 
Since $H_{11}^{chd}$ is the hypothesis that exhibits the most substantial evidence over the null hypothesis (this hypothesis was supported in Ref.~\cite{vendrell_2003} with $p=0.0056$), we will take $H_{11}^{chd}$
as our working hypothesis upon which we will base our subsequent inferences.

\begin{table}[t]
\begin{tabular}{c|ccccc
}
$B_{row,col}
$ ~&~ ${H_{00}}$~&~${H_{01}}$~&~${H_{10}}$~&~${H_{11}}$~&~${H_{11}^{chd}}$
\\ 
\hline
$H_{00}$&0 &  &  &  & 
\\ 
$H_{01}$&3.99  &0 & &&
\\ 
$H_{10}$&4.37  &0.91 &0 &&
\\ 
$H_{11}$&4.08  &1.02 &0.93 &0 &
\\ 
$H_{11}^{chd}$&7.07 &1.77 &1.62 &1.73 & 0
\\ \hline
\end{tabular}
\caption{\label{table:dyo} {\bf The Bayes factors for the hypotheses considered.} Here $B_{row,col}=H_{row}/H_{col}.$}
\end{table}

\subsection{Model fitting}

Having inferred from the computation of the evidence which of the
possible hypotheses is most likely to be compatible with the data in the presence of the SNP, we proceed to compute the 
probability for the occurrence of the polymorphism. 
Let $H$ denote our working hypothesis. Then the probability that the SNP
is present is
\ba
P(SNP\vert H)
&=&\int d\theta~P(SNP\vert \theta)P(\theta\vert D\wedge H)
=\int d\theta~P(SNP\vert \theta)
  { {P( D\vert \theta \wedge H)P(\theta\vert H)}\over{P(D\vert H)}}.\quad
\ea
Since $H_{11}^{chd}$ consists of a two-component hypothesis \cite{fischer_2009}, each of which described by two parameters,  the resulting parameter space is four-dimensional. The equation above must be generalized for a multidimensional parameter space where each factor is no longer a scalar but instead a (4x4) matrix. Since each component consists of two disjoint sets, the matrix is diagonal,
each component being weighted by the relative size of the population. 
We then write
\ba
P(SNP\vert H)
=\int d^2p~
P(SNP\vert p)
P( p\vert D\wedge H)
+\int d^2\bar p~
P(SNP\vert \bar p)
P( \bar p\vert D\wedge H).
\ea
Here the indices range over the two-dimensional parameter spaces, with
$p=(p_{01},p_{11})$ and $\bar p=(p_{0\bar 1},p_{\bar 1 1}),$ where for simplicity we have dropped the tilde from the notation used in Appendix \ref{app:evidence}. 
In particular, $p_{01}$ is the frequency of SNP given
the occurrence of CHD 
and $p_{0\bar 1}$ the frequency of SNP given non-occurrence of CHD, both 
subject to non-occurrence of t2DM, whereas
$p_{11}$ is the frequency of SNP given that t2DM has
occurred and $p_{\bar 1 1}$ the frequency of SNP given that t2DM has not occurred, both subject to CHD having occurred.
Let
$P(SNP\vert p)=p$ and $P(SNP\vert \bar p)=\bar p.$ 
The posterior probability of $p$ is by the Bayes theorem
\ba
P( p\vert D\wedge H)
&=&{ {P( D\vert p\wedge H)P(p\vert H)}
 \over {P(D\vert H)} }
={ P( D\vert p\wedge H)
 \over {P(D\vert H)} }
\ea
and similarly for the posterior probability of $\bar p.$
In the last step we assume for simplicity a uniform prior for both $p$ and
$\bar p.$\footnote{This choice of prior is justified by the 
absence of an {\it a priori} bias on the values of these parameters.}
Writing the evidence as
\ba
P(D\vert H)
=\int d^2p~ P(D\vert p\wedge H)P(p\vert H)
+\int d^2\bar p~P(D\vert \bar p\wedge H)P(\bar p\vert H)
\ea
we find for hypothesis $H_{11}^{chd}$ that
\ba
&&P(SNP\vert H)=\cr
&=&{1\over {P(D\vert H)}}
\int~dp_{01}\int dp_{11}\int dp_{0\bar 1}\int dp_{1 \bar 1}\cr
&&\times \Biggl[ \gamma{N_{CHD,\overline{t2DN}} \choose N_{SNP,CHD,\overline{t2DN}}}
 p_{01}^{N_{SNP,CHD,\overline{t2DN}}+1}(1-p_{01})^{N_{\overline{SNP},CHD,\overline{t2DN}}}\cr
&&\quad+(1-\gamma){N_{CHD,t2DN} \choose N_{SNP,CHD,t2DN}}
 p_{11}^{N_{SNP,CHD,t2DN}+1}(1-p_{11})^{N_{\overline{SNP},CHD,t2DN}}
\cr
&&\quad+
\tilde\gamma
 {N_{\overline{CHD},\overline{t2DN}}\choose N_{SNP,\overline{CHD},\overline{t2DN}}}
  p_{0\bar 1}^{N_{SNP,\overline{CHD},\overline{t2DN}}+1}
   (1-p_{0\bar 1})^{N_{\overline{SNP},\overline{CHD},\overline{t2DN}}}\cr
&&\quad
+(1-\tilde \gamma)
 {N_{CHD,\overline{t2DN}} \choose N_{SNP,CHD,\overline{t2DN}}}
  p_{\bar 1 1}^{N_{SNP,CHD,\overline{t2DN}}+1}
   (1-p_{\bar 1 1})^{N_{\overline{SNP},CHD,\overline{t2DN}} }\Biggr]
\ea
which yields
\ba
P(SNP\vert H)
&=&{1\over P(D\vert H)}\cr
&&\times
\Biggl[
\gamma
 {{N_{SNP,CHD,\overline{t2DN}}+1}
  \over { (N_{CHD,\overline{t2DN}}+2) (N_{CHD,\overline{t2DN}}+1)}}\cr
&&\quad+(1-\gamma)
 {{N_{SNP,CHD,t2DN}+1}
  \over { (N_{CHD,t2DM}+2)(N_{CHD,t2DM}+1)}}
  \Biggr]\cr
&&\quad +
\tilde \gamma
 {{N_{SNP,\overline{CHD},\overline{t2DN}}+1}
  \over { (N_{\overline{CHD},\overline{t2DN}}+2) 
    (N_{\overline{CHD},\overline{t2DN}}+1)}}\cr
&&\quad+(1-\tilde \gamma)
 {{N_{SNP,CHD,\overline{t2DN}}+1}
  \over { (N_{CHD,\overline{t2DM}}+2)(N_{CHD,\overline{t2DM}}+1)}}
  \Biggr].
\label{eqn:P_SNP}
\ea
Here $\gamma=N_{CHD,\overline{t2DM}}/N_{CHD}$ and 
$\tilde \gamma=N_{\overline{CHD},\overline{t2DM}}/N_{\overline{t2DM}},$ 
with $P(D\vert H)$ given by Eqn.~(\ref{eqn:h_11^chd}).
Substituting the values from Table~\ref{table:ena} 
we find that $P(SNP\vert H)=0.30\pm 0.001,$
which we identify as the effective mutation rate $\lambda_{eff}$ of the
Poisson probability distribution describing the occurrence of the
SNP. Although comparable, this value is different from the na\"ive guess 
$\lambda=N_{SNP}/N=0.27$ or from the more elaborate one arising 
from the assumption of the null hypothesis [see Appendix \ref{h00} for the derivation].
It then follows that the posterior probability for the occurrence of
$n$ mutations in a population of size $N$ is  
\ba
P(n\vert \lambda_{eff}\wedge N)
=\exp[-\lambda_{eff}N]{(\lambda_{eff}N)^n\over {n!}}.
\ea
Similarly we compute
\ba
P(SNP\vert CHD\wedge H)
&=&\int d^2p~P(SNP\vert CHD\wedge p)P(p\vert D\wedge H)
=\int d^2p~{ {p~P(D\vert p\wedge H)}\over P(D\vert H)}\cr
&=&{1\over P(D\vert H)}\Biggl[
\gamma
 {{N_{SNP,CHD,\overline{t2DN}}+1}
  \over { (N_{CHD,\overline{t2DN}}+2) (N_{CHD,\overline{t2DN}}+1)}}\cr
&&+(1-\gamma)
 {{N_{SNP,CHD,t2DN}+1}
  \over { (N_{CHD,t2DM}+2)(N_{CHD,t2DM}+1)}}
  \Biggr]
\label{eqn:P_SNP|CHD}
\ea
and find that $P(SNP\vert CHD\wedge H)=0.20\pm 0.001.$
We can also compute
\ba
P(t2DN \vert CHD\wedge H)
&=&\int dp~P(t2DM\vert CHD\wedge p)P(p\vert D\wedge H)\cr
&=&{1\over P(D\vert H)}{ {N_{CHD,t2DN}+1}\over { (N_{CHD}+2)(N_{CHD}+1)}}
\label{eqn:P_t2DM|CHD}
\ea
finding that 
$P(t2DN \vert CHD\wedge H)=0.09\pm 0.001.$
The errors indicated were computed from error propagation, assuming the error of a counting result $n$ to be $1/\sqrt{n}.$

\section{Top-down approach}
\label{section:top-down}

We now proceed to estimate the influence of the SNP on the
occurrence of CHD. We want to find the posterior probability of the
occurrence of CHD given the presence of the SNP, i.e.
\ba
P(CHD\vert SNP\wedge H)
={P(SNP\vert CHD\wedge H)P(CHD\vert H)\over P(SNP\vert H)}. 
\label{eqn:P_CHD|SNP}
\ea
Here $P(CHD\vert H)$ is the prior probability of CHD and
$P(SNP\vert CHD \wedge H)$ is the likelihood of CHD for a fixed SNP.
The remaining term $P(SNP\vert H)$ has no CHD dependence and can thus
be absorbed into the normalization constant. It is known as the
evidence or the marginal likelihood. 

\subsection{A simplistic model for the signalling pathway}

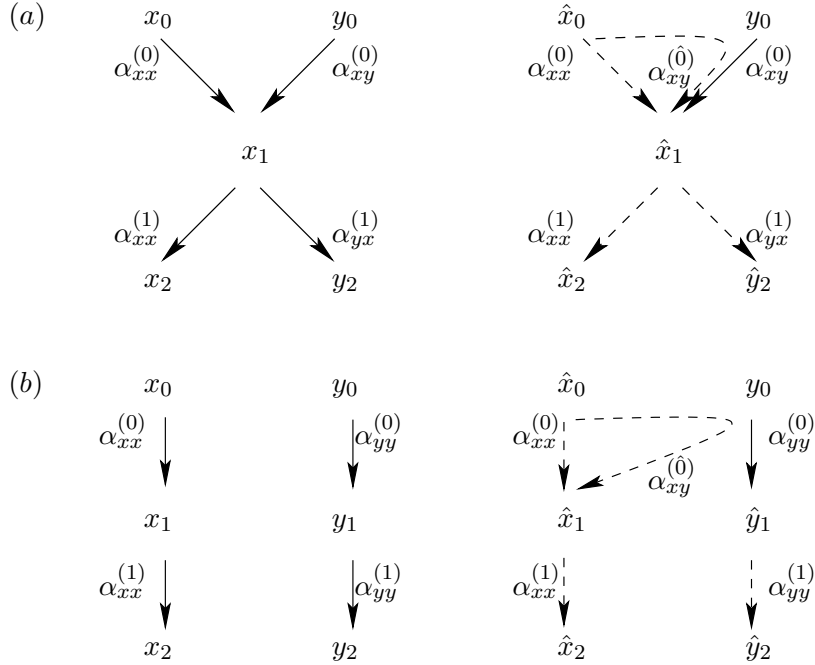
\begin{figure}[t]
\setlength{\unitlength}{1cm}
\hspace{2.cm}
\includegraphics[width=8.cm]{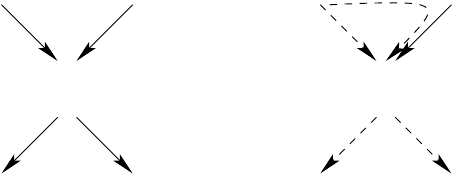}
\put(-10,3.2){$(a)$}
\put(-8.2,3.2){$x_{0}$}
\put(-5.7,3.2){$y_{0}$}
\put(-8.6,2.5){$\alpha_{xx}^{(0)}$}
\put(-5.7,2.5){$\alpha_{xy}^{(0)}$}
\put(-6.9,1.4){$x_{1}$}
\put(-8.6,0.3){$\alpha_{xx}^{(1)}$}
\put(-5.7,0.3){$\alpha_{yx}^{(1)}$}
\put(-8.2,-0.3){$x_{2}$}
\put(-5.7,-0.3){$y_{2}$}
\put(-2.7,3.2){$\hat x_{0}$}
\put(-0.2,3.2){$y_{0}$}
\put(-3.1,2.5){$\alpha_{xx}^{(0)}$} 
\put(-0.2,2.5){$\alpha_{xy}^{(0)}$}
\put(-1.5,2.45){$\alpha_{xy}^{(\hat 0)}$}
\put(-1.4,1.4){$\hat x_{1}$}
\put(-3.1,0.3){$\alpha_{xx}^{(1)}$} 
\put(-0.2,0.3){$\alpha_{yx}^{(1)}$}
\put(-2.7,-0.3){$\hat x_{2}$}
\put(-0.2,-0.3){$\hat y_{2}$}
\vfill
\vspace{1.cm}
\hspace{2.cm}
\includegraphics[width=8.0cm]{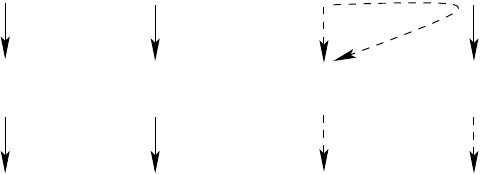}
\put(-10,3.2){$(b)$}
\put(-8.2,3.2){$x_{0}$}
\put(-5.7,3.2){$y_{0}$}
\put(-8.8,2.5){$\alpha_{xx}^{(0)}$}
\put(-5.4,2.5){$\alpha_{yy}^{(0)}$}
\put(-8.2,1.4){$x_{1}$}
\put(-5.7,1.4){$y_{1}$}
\put(-8.8,0.5){$\alpha_{xx}^{(1)}$}
\put(-5.4,0.5){$\alpha_{yy}^{(1)}$}
\put(-8.2,-0.3){$x_{2}$}
\put(-5.7,-0.3){$y_{2}$}
\put(-2.7,3.2){$\hat x_{0}$}
\put(-0.2,3.2){$y_{0}$}
\put(-3.3,2.5){$\alpha_{xx}^{(0)}$}
\put(-1.5,1.9){$\alpha_{xy}^{(\hat 0)}$}
\put(0.1,2.5){$\alpha_{yy}^{(0)}$}
\put(-2.7,1.4){$\hat x_{1}$}
\put(-0.2,1.4){$\hat y_{1}$}
\put(-3.3,0.5){$\alpha_{xx}^{(1)}$}
\put(0.1,0.5){$\alpha_{yy}^{(1)}$}
\put(-2.7,-0.3){$\hat x_{2}$}
\put(-0.2,-0.3){$\hat y_{2}$}
\caption{\small \baselineskip=0.5cm {
{\bf Diagram of the pathways with different source signals. }
The first column depicts the original architecture corresponding to the canonical pathway (i.e., in the absence of the SNP). The second column depicts one possible alteration of the corresponding pathway due to the SNP. Here the SNP acts on the signal $x_0.$ The continuous lines represent the functioning pathways, and the dashed lines represent pathways which can suffer delay or cease to function entirely (in the case of previously existing pathways), or which could simply not come to exist (in the case of newly generated pathways). 
(a) Case 1a). For $y_0=0$ this reduces to Case 2. (b) Case 1b). }}
\label{fig:pikassoyla2}
\end{figure}

Since the working hypothesis relates the presence of the SNP with
both the occurrence of CHD and the occurrence of t2DM, 
we infer that 
the changes from the canonical pathway 
introduced by the SNP will have repercussions on the signalling cascades which
regulate the emergence of CHD and t2DM. 
If we assume that the SNP will only affect one source signal,
then this correlation suggests that the resulting signal transduction
pathways interfere with one another.
Functional interference can be derived from a common source signal or
from common components downstream \cite{komarova_2005,komarova_2007}. 
As far as the source signal is concerned,
we allow for two possibilities:
1) the pathways have different source signals;
2) the pathways have the same source signal and sufficiently
downstream diverge.
In case 1) and in order to reproduce interference between the two signalling
pathways we can still have two further sub-cases:
1a) the pathways share components and the effect of the SNP consists
of either an alteration in the velocity of the affected signal or an alteration
of one of the pathways;
1b) the pathways do not share components but the pathway altered by the SNP
shares components with the unaltered one. 
These two subcases can be distinguished by the correlation between the
two diseases in the absence of the SNP, with case 1a) describing the
existence of an {\it a priori} correlation between the occurrence of
the two diseases and case 1b) the absence of such correlation. 
In case 2) the interference is built-in so the effect of the SNP
is similar to that in case 1a). 
The diagrams are depicted in Fig.~\ref{fig:pikassoyla2} which we proceed now to describe.

The variables $x$ and $y$ describe {\it black boxes} along the
pathways that regulate the emergence of CHD and t2DM
respectively. By black boxes we mean unresolved chemical reactions
where no intervening elements are specified other than the input and the
output reaction rates between two sequential black boxes. 
The index of the variables denotes the relative position in the
pathway of the corresponding black box component. 
Thus '$0$' indicates the upstream component fed by the
initial signal and which is subject to alteration
upon the action of the SNP, whereas '$2$' indicates the final component
which determines the emergence of the disease, with '$1$' denoting the
intermediary component where the interference of the altered pathway
with the unaltered one is manifested. Here the SNP affects the signal
$x_0$ and propagates downstream through an altered or newly created
pathway '$\hat 0$'. A variable which refers to an altered pathway is denoted by $\hat x_{0},$ whereas a variable which refers to a newly created pathway is denoted by $x_{\hat 0}.$ The same rule holds for the corresponding rates.
The cases where the original pathway ceases to function can be interpreted as inhibition. This a simplistic description of the potential signalling pathways involved, which is but a caricature of the real biochemical system. Nonetheless it is the possible description based on the data, which moreover captures the functional correlations inferred.

To select the working model we use as criterion the possibility of the emergence of
a connection between the SNP and each disease without the participation of the other.
We conceive three possible ways for the SNP to function: 
(1) the signal does not suffer any alteration from that in the absence of the SNP;
(2) the signal is triggered at a smaller rate so that 
 $\hat x_{0}<x_{0};$
(3) no signal is triggered. We can exclude (1) on the basis that it contradicts the working hypothesis. 
Moreover, since (3) can be described as a limiting case of (2) when 
$\hat x_{0}=0,$ we proceed to solely analyse (2)
with each suggested model according to the disease combinations which can be reproduced.

\paragraph*{Case 1a)}
If both $x_{2}$ and $y_{2}$ require that $x_{1}$ be acted by the products of both $x_{0}$ and $y_{0},$ then the 
functional time scales 
of the two pathways should be very close. 
In the case of a slower reaction rate, if the signalling pathways are assumed to consist of an isolated system then both diseases would occur. The way to prevent it would be by capturing the required reagents from neighbouring pathways. 
This scenario, however, would be beyond our current capabilities of inference and constraint. 
If instead $x_{1}$ has the flexibility of being independently activated by the two signals and of also independently acting upon $x_{2}$ and $y_{2},$ then  
the emergence of the diseases will depend on the supply of each pathway that develops from $x_{1}$ downstream. Should a single triggering signal be enough to supply for both downstream pathways, it could be the case that no disease  
occurs. This would moreover depend on the phase difference between the generation of $x_{1}$ upon the activation of $x_{0}$ and $y_{0}.$ Should one triggering signal not be enough, then either disease could occur. This could be prevented if a compensation mechanism were triggered so that, in the absence of an effective signalling from one source, 
the working source would be stimulated according to the deficiency.



\paragraph*{ Case 1b)}
In this case, $x_{1}$ and $y_{1}$ are independently sourced and independently develop their pathways downstream. In the case of a slower reaction rate, no interference would be generated as long as the SNP-affected pathway could still run. However, in the case of a more drastic reduction of the trigger of $x_{1},$ an alternative pathway would be intercepted and the required reagents deviated. This could lead to a case of competition. Should pathway $x$ be thus maintained, pathway $y$ could either collapse or continue, resulting in the disease implicated in the pathway $y$ emerging or not. If pathway $x$ cannot be maintained and the corresponding disease is not avoidable, then we can still have either emergence or not of the other disease, depending on the degree of reagent deviation.

This empirical analysis can be complemented by the following quantitative one.
We will assume that the system under study can be described as a dynamical one.
In the absence of the SNP, the dynamical system is described by the rates indicated in Fig.~\ref{fig:pikassoyla2}, first column.
If the SNP has occurred in the coding of the input signal in $x_{0},$ then the system will instead be sourced by the altered $\hat x_{0}$ and described by the rates as indicated in Fig.~\ref{fig:pikassoyla2}, second column.

We compute the fixed points (identified by the superscript '$*$') of the quantities involved in both the canonical pathway and the SNP--altered one. 
The fixed points describe the state of the system in dynamical equilibrium and are computed by setting the time derivates equal to zero. Since we are interested in the probabilities which describe the average states of the system and not in the dynamical evolution that leads to those states, the fixed points are the variables to be used.


For case 1a) 
\ba
{d\hat x_{1}\over dt}&=&
\alpha_{xx}^{(0)}\hat x_{0}
+\left[ \alpha_{xy}^{(0)}
+\alpha_{xy}^{(\hat 0)}\beta(\hat x_{0})\right]
y_{0}
-\left( \alpha_{xx}^{(1)} +\alpha_{yx}^{(1)}
+\delta_{x}^{(1)}\right)\hat x_{1} \\
{d\hat x_{2}\over dt}&=&
 \alpha_{xx}^{(1)}\hat x_{1} 
 -\delta_{x}^{(2)}\hat x_{2} \\
{d\hat y_{2}\over dt}&=&
 \alpha_{yx}^{(1)}\hat x_{1} -\delta_{y}^{(2)}\hat y_{2}
\ea
where the $\delta$'s denote decay rates.
Here $\beta$ is the fraction of the output of $y_0$ which activates $x_{1}$
and thereby attempts to compensate for the deficiency of activation of $x_1$ derived from the SNP
\ba
\label{eqn:beta}
\beta(\hat x_{0})=
\left( 1-{\hat x_{0}\over {x_{0}-\Delta_{0}^{thres}}}\right)
 \Theta[x_{0}-\Delta_{0}^{thres}-\hat x_{0}]
\ea
where $\Theta$ is the Heaviside step function.
This term is reminiscent of the carrying capacity term which sets an upper limit to cellular growth \cite{komarova}. Here we set a lower limit, as determined by $\Delta_{0}^{thres},$ to the deficiency in $x_{0}$ caused by the altered signal for the onset, as imposed by the step function, of the compensating mechanism. This quantity would measure a functional marker for the presence of the SNP.
We compare the fixed points in both the canonical pathway and the SNP-altered one, finding that
\ba
\hat x_{2}^{*}-x_{2}^{*}
&=&\left[ \alpha_{xx}^{(0)}
\left( \hat x_{0}-x_{0}\right)
+\alpha_{xy}^{(\hat 0)}\beta(\hat x_{0})y_{0}
\right]
{\alpha_{xx}^{(1)}\over {\delta_{x}^{(2)}} }
  {1\over {\alpha_{xx}^{(1)}+\alpha_{yx}^{(1)}+\delta_{x}^{(1)}} }
\label{eqn:delta2_1a}\\
\hat y_{2}^{*}-y_{2}^{*}
&=&\left[ \alpha_{xx}^{(0)}
\left( \hat x_{0}-x_{0}\right)
+\alpha_{xy}^{(\hat 0)}\beta(\hat x_{0})y_{0}
\right]
{\alpha_{yx}^{(1)}\over {\delta_{y}^{(2)}} }
  {1\over {\alpha_{xx}^{(1)}+\alpha_{yx}^{(1)}+\delta_{x}^{(1)}} }.
\ea
This shows that, in the absence of the compensating term in $y_{0},$ a difference in $x_{0}$ will be reflected in a difference in both $x_{2}^{*}$ and $y_{2}^{*},$ and thus imply the occurrence respectively of CHD and diabetes. 
Thus the occurrence of each disease will be related to both the onset of the altered signal and to the change from the canonical pathway of its propagation downstream. 
Expressing $\hat x_{0}$ and $\hat y_{2}^{*}$ as functions of $\hat x_{2}^{*},$ we find that 
\ba
\hat x_{0}&=&
{1\over A_{0}} \left( A_{2}\hat x_{2}^{*}-A\right)\\
\hat y_{2}^{*}&=&{\hat x_{2}^{*}\over B_{2}}
\ea
where the $A$'s and $B$'s are functions of the biochemical rates, the canonical signal $x_{0}$ and the fixed point of the canonical pathway $x_{2}^{*}.$ 


For case 1b)
\ba
{d\hat x_{1}\over dt}&=&
\alpha_{xx}^{(0)}\hat x_{0}
+
\alpha_{xy}^{(\hat 0)}\beta(\hat x_{0})y_{0}
-\left( \alpha_{xx}^{(1)}
+\delta_{x}^{(1)}\right)\hat x_{1} \\
{d\hat y_{1}\over dt}&=&
\alpha_{yy}^{(0)}
  \left[ 1-\beta(\hat x_{0})
  \right]y_{0}
-\left(\alpha_{yy}^{(1)}
+\delta_{y}^{(1)}\right)\hat y_{1} \\
{d\hat x_{2}\over dt}&=&
 \alpha_{xx}^{(1)}\hat x_{1} 
 -\delta_{x}^{(2)}\hat x_{2} \\
{d\hat y_{2}\over dt}&=&
 \alpha_{yy}^{(1)}\hat y_{1} -\delta_{y}^{(2)}\hat y_{2}.
\ea
The fixed points for the canonical and altered pathways are related as follows
\ba
\hat x_{2}^{*}-x_{2}^{*}
&=&\left[ \alpha_{xx}^{(0)}
\left( \hat x_{0}-x_{0}\right)
+\alpha_{xy}^{(\hat 0)}\beta(\hat x_{0})y_{0}
\right]
{\alpha_{xx}^{(1)}\over {\delta_{x}^{(2)}} }
  {1\over {\alpha_{xx}^{(1)}+\delta_{x}^{(1)}} }
\label{eqn:delta2_1b}\\
\hat y_{2}^{*}-y_{2}^{*}
&=&-\alpha_{yy}^{(\hat 0)}\beta(\hat x_{0})y_{0}~
{\alpha_{yx}^{(1)}\over {\delta_{y}^{(2)}} }
  {1\over {\alpha_{yy}^{(1)}+\delta_{y}^{(1)}} }.
\ea
In this case, the change from the sharing of a component while compensating for the supply of one pathway could cause the other to be depleted of essential reagents. The outcome, however, will depend on how much the intercepting pathway takes and how much the intercepted pathway can run canonically without.
Similarly we find expressions for $\hat x_{0}$ and $\hat y_{2}^{*}$ as functions of $\hat x_{2}^{*}$  
\ba
\hat x_{0}&=&
{1\over A_{0}} \left( A_{2}\hat x_{2}^{*}-A\right)\\
\hat y_{2}^{*}&=&{1\over B_{2}}
 \left[ \left( A_{2}\hat x_{2}^{*}-A\right){B_{0}\over A_{0}}+B\right]
\ea
where similarly the $A$'s and $B$'s are functions of the biochemical rates, the canonical signal $x_{0}$ and the fixed points of the canonical pathways $x_{2}^{*}$ and $y_{2}^{*}.$

In the following subsection we will use as the working description of the signalling pathway the dynamical system of case 1a). Given the similarity in the functional form of the dependence of the variables, similar conclusions would also be inferred, with the interpretation only differing on the basis of the different structure of the pathways. 
 
 \subsection{Translation of the dynamical system into a probability description}

In order to compute $P(CHD\vert SNP\wedge H),$ we proceed to write the probabilities in Eqn.~(\ref{eqn:P_CHD|SNP}) in terms of the variables in the description of the dynamical system. 
In particular, we want to compute
\ba
P(x_{2}\vert x_{0})={ {P(x_{0}\vert x_{2})P(x_{2})}\over {P(x_{0})} }
\ea
where the 
probability $P(x_{2})$ is related to that for the occurrence of CHD and the likelihood $P(x_{0}\vert x_{2})$ is related to that for the data on the SNP conditional on the prior for CHD. The quantity $P(x_{0})$ is the evidence, which 
is found by marginalising the likelihood and is related to the probability for occurrence of the SNP.
We will use the probabilities computed in the previous
section to constrain the priors assumed here. We will then derive an expression for $P(CHD\vert SNP\wedge H)$ in terms of both the statistical properties of the priors and the biochemical parameters of the transmission process from the SNP to the CHD. 

We assume that the 
probability for the occurrence of CHD is described by a Gaussian distribution with expectation value equal to the fixed point of the final component of the pathway 
 $\mu_{2_P}=x_{2}^{*},$ and standard deviation $\sigma_{2_P}$
\ba
P(x_2)={1\over {\sqrt{2\pi} \sigma_{2_P} }}~
 \exp \left[ -{ \left( x_{2}-\mu_{2_P}\right)^2 \over {2\sigma_{2_P} ^2}}\right].
\ea
In the probability description, the quantities $\mu_{2_P}$ and $\sigma_{2_P}$ are properties of the prior knowledge of the distribution of the occurrence of CHD derived from population sampling and expressed in terms of the biochemical parameters of the system according to the model considered. In the absence of further data, these quantities characterize a theoretical prior which we can assume to be approximated by a binomial distribution for sufficiently large $N_{CHD}.$ The parameter of the binomial distribution 
is the frequency of occurrence of CHD which has for maximal likelihood estimator 
$p_{CHD}=N_{CHD}/(N_{CHD}+N_{\overline{CHD}}).$
The mean and the variance of the approximated Gaussian distribution are given by
\ba
\mu_{2_P}=N_{CHD}~p_{CHD},\quad
\sigma_{2_P}^2=N_{CHD}~p_{CHD}(1-p_{CHD})
\ea
which yield respectively $\mu_{2_P}=170\pm1$ and $\sigma_{2_P}=9.2\pm1.6.$ 
The fixed point corresponding to the canonical pathway denotes absence of disease, 
whereas deviations from this value will entail a non-vanishing probability of occurrence of CHD. 
In order to quantify this probability, we need to devise a criterion to determine the emergence of CHD.
Deviations on the fixed point of an altered pathway from that of the canonical pathway are quantified by 
$\Delta_{2}=\mu_{2_P}- \hat x_{2}^{*}.$\footnote{
In accordance with the description encapsulated in the 
model, the SNP will act by causing deficiency in the {\it modus operandi} of the system. Should it instead act by causing excess, then 
a symmetric interval about $\mu_{2_P}$ would be the generalization to account for possible saturation and consequent screening effect. 
The changes to implement in all the subsequent results would be straightforward.}
This quantity would measure a non-environmental marker\footnote{The distinction between functional and environmental markers can be shady and will thus require care.}
for the occurrence of CHD \cite{robinson_2006}.
For deviations larger than a threshold value $\Delta_2^{thres}$ the disease will occur.
The probability of occurrence of CHD will be
\ba
P(CHD)
&=&P(x_2<\mu_{2_P}-\Delta_2^{thres})\cr
&=&{1\over {\sqrt{2\pi} \sigma_{2_P}}}
\int _{-\infty}^{\mu_{2_P}-\Delta_2^{thres}}dx_{2}~
  \exp \left[ -{ \left( x_{2}-\mu_{2_P}\right)^2 \over {2\sigma_{2_P} ^2}}\right].
   \qquad
\ea
Deviations at the level of $x_{2}$ will be the result of the propagation along the pathway of deviations at the level of $x_{0}$ according to Eqn.~(\ref{eqn:delta2_1a}) or (\ref{eqn:delta2_1b}), depending on the model considered.

The likelihood of the data on $x_0$ given the variable $x_2$ 
is assumed also to follow a Gaussian distribution centred at $\hat x_{0}$ and with standard deviation $\sigma _{0_D}$ \cite{lupton}
\ba
P(\hat x_{0}\vert \hat x_{2}^{*})
={1\over {\sqrt{2\pi} \sigma_{0_D}}}~
 \exp \left[ -{ \left( \hat x_{0}-\hat x_{2}^{*}\right)^2 \over {2\sigma_{0_D} ^2}}\right].
\ea
Here the quantities $\hat x_{0}$ and $\sigma_{0_D}$ describe properties of the data in the presence of the SNP expressed in terms of the biochemical parameters of the system.
The likelihood of CHD will be given by the integral in $x_{2}$ because the SNP enters as data through the modelling of the system.
Substituting $\hat x_{0}=\hat x_{0}(\hat x_{2}^{*})=(A_2~\hat x_{2}^{*}-A)/A_0,$ where the $A$'s are functions of the statistical parameters $\mu_{2_P},$ $x_{0}$ and $\sigma_{0_D}$ as well as of the biochemical parameters, we find that
\ba
P(SNP\vert CHD) 
&=&{1\over {\sqrt{2\pi} \sigma_{0_D}}}
\int _{-\infty}^{\mu_{2_P}-\Delta_2^{thres}}d\hat x_{2}^{*}~
  \exp \left[ -{ \left( \hat x_{0}-\hat x_{2}^{*}\right)^2 \over {2\sigma_{0_D} ^2}}\right] \cr
&=&{1\over 2}
+{A_{0}\over {2(A_{0}-A_{2})}}~
{\rm erf}\left[ 
 {{A+(\mu_{2_P}-\Delta_2^{thres})(A_0-A_2)}\over
   {\sqrt{2}\sigma_{0_D}}A_0}\right]\quad
\ea
where {\it erf} stands for the error function given by the integral ${\rm erf}(x)=(2/\sqrt{\pi})\int_{0}^{x}dy~\exp[-y^2].$ This probability was computed in Eqn.~(\ref{eqn:P_SNP|CHD}). 

We can now derive the functional form of the evidence in terms of the statistical and the biochemical parameters.
Combining the two assumptions above, we find that
\ba
 {P(\hat x_0\vert \hat x_2^{*})P(\hat x_2^{*})}
={1\over {2\pi \sigma_{0_D}\sigma_{2_P}}}~
   \exp\left[ -{1\over 2}\left( \Gamma 
   -\mu_{eff}^2\right)\right]
     \exp\left[ -{ {\left( \hat x_2^{*} 
     -\mu_{eff}\right)^2}\over {2\sigma_{eff}^2} }\right]\quad
\ea
where 
\ba
&&\mu_{eff}
=\left( {\mu_{2_P}\over \sigma_{2_P}^2}
-{\hat x_{0}\over \sigma_{0_D}^2}\right)\bigg/
  \left( {1\over \sigma_{2_P}^2} +{1\over \sigma_{0_D}^2}\right), \\
&&\sigma_{eff}^2
=1\bigg/ \left( {1\over \sigma_{2_P}^2} +{1\over \sigma_{0_D}^2}\right),\\
&&\Gamma=\left( {\mu_{2_P}^2\over \sigma_{2_P}^2}
 +{\hat x_{0}^2\over \sigma_{0_D}^2}\right). ~~
\ea
Substituting $\hat x_{0}=\hat x_{0}(\hat x_{2}^{*}),$ we integrate in 
$\hat x_{2}^{*}$ finding 
that
\ba
&&P(SNP)
=\int ^{+\infty}_{-\infty}
 d\hat x_{2}^{*}~P(\hat x_{0}\vert \hat x_{2}^{*})~P(\hat x_{2}^{*})\cr
&=&{1\over \sqrt{2\pi}}
{1\over \sqrt{\sigma_{0_D}^2 +\sigma_{2_P}^2(A_{0}-A_{2})^2/A_{0}^2} }~
 \exp\left[
-{1\over 2}
{ {[A/A_{0}+\mu_{2_P}(A_{0}-A_{2})/A_{0}]^2}\over
{\sigma_{0_D}^2+\sigma_{2_P}^2(A_{0}-A_{2})^2/A_{0}^2}
}
\right].\quad\quad~~
\label{eqn:P_x0}
\ea
If we furthermore assume that $x_{0}$ follows a Gaussian distribution centred at the value for the canonical path, which we denote by $\mu_{0},$ and with standard deviation $\sigma_{0_D},$ which is such that 
$\Delta_{0}^{thres}=\sqrt{2\sigma_{0_D}^2\ln[1/P(\hat x_0)]}$ when 
$\hat x_{0}=\mu_0-\Delta_{0}^{thres},$ 
then following a similar reasoning to that for CHD, we will have presence of the SNP for 
$\hat x_{0}<\mu_{0}-\Delta_{0}^{thres}.$ 
Hence
\ba
P(SNP)&=&
{1\over {\sqrt{2\pi}\sigma_{0_D}}}
 \int _{-\infty}^{\mu_{0}-\Delta_{0}^{thres}}d\hat x_{0}~
  \exp\left[-{(\hat x_{0}-\mu_{0})^2\over {2\sigma_{0_D}^2}}\right]
={1\over 2}
-{1\over 2}~{\rm erf}\left[ 
{\Delta_{0}^{thres}\over {\sqrt{2}\sigma_{0_D}}}\right]
\ea
which equals $\lambda_{eff}=0.30\pm 0.001,$ as computed in Eqn.~(\ref{eqn:P_SNP}), and thus serves to constrain the parameters in Eqn.~(\ref{eqn:P_x0}).
We can also solve for $\Delta_{0}^{thres}$ finding that 
$\Delta_{0}^{thres}=(0.52\pm0.006)\sigma_{0_D}.$ The quantity $\Delta_{0}^{thres}$ determines the parameter $\beta$ in Eqn.~(\ref{eqn:beta}).
Combining the two conditions above, we find the value for 
$P(\hat x_{0})=0.87\pm0.003$ which we can interpret as the probability that the SNP has occurred when $\hat x_{0}$ is below the threshold value that can trigger the canonical pathway. 

Moreover, having in Eqn.~(\ref{eqn:P_t2DM|CHD}) also computed 
$P(t2DM\vert CHD),$ we write the corresponding likelihood 
\ba
P(\hat y_{2}^{*}\vert x_{2}^{*})
={1\over {\sqrt{2\pi} s_{2_D}}}~
 \exp \left[ -{ \left( \hat y_{2}^{*}-\hat x_{2}^{*}\right)^2 \over {2s_{2_D} ^2}}\right].
\ea
Note that we cannot follow a reasoning analogous to that for the case of the probability of $CHD$ because the population $N_{t2DM}$ is not entirely random.
Substituting $\hat y_{2}^{*}=\hat y_{2}^{*}(\hat x_{2}^{*})=\hat x_{2}^{*}/B_{2},$ where $B_{2}$ is a function of the biochemical parameters,
we find that
\ba
P(t2DM\vert CHD)
&=&
{1\over {\sqrt{2\pi} s_{2_D}}}
 \int _{-\infty}^{\mu_{2_P}-\Delta_2^{thres}}d\hat x_{2}^{*}~
  \exp \left[ -{ \left( \hat y_{2}^{*}-\hat x_{2}^{*}\right)^2 \over {2s_{2_D} ^2}}\right]\cr
&=&{1\over 2}
+{B_{2}\over {2(1-B_2)}}~
{\rm erf}\left[
 { { (\mu_{2_P}-\Delta_{2}^{thres})(1-B_{2})}\over {\sqrt{2}s_{2_D}B_{2}} }
\right].
\ea

We can now compute the posterior probability of the variable $x_2$ given 
$\hat x_0,$ i.e. 
\ba
P(\hat x_2^{*}\vert \hat x_0)
={1\over {2\pi \sigma_{0_D}\sigma_{2_P}}}{1\over P(\hat x_{0})}~
   \exp\left[ -{1\over 2}\left( \Gamma 
   -\mu_{eff}^2\right)\right]
     \exp\left[ -{ {\left( \hat x_2^{*} 
     -\mu_{eff}\right)^2}\over {2\sigma_{eff}^2} }\right],
\quad\quad
\ea
finding for the probability that CHD will occur given that SPN has occurred that
\ba
&&P(CHD\vert SNP)
=
{1\over P(SNP)}
 \int _{-\infty}^{\mu_{2_P}-\Delta_{2}^{thres}}
 d\hat x_{2}^{*}~P(\hat x_0\vert \hat x_2^{*})~P(\hat x_2^{*})\cr
&=&{1\over 2}
+{1\over 2}~{\rm erf}\left[
{ {-\sigma_{0_D}^2\Delta_2^{thres} +
\sigma_{2_P}^2[A+(\mu_{2_P}-\Delta_{2})(A_{0}-A_{2})](A_{0}-A_{2})/A_{0}^2}
 \over {\sqrt{2} \sigma_{0_D}\sigma_{2_P}
  \sqrt{\sigma_{0_P}^2+\sigma_{2_P}^2(A_{0}-A_{2})^2/A_{0}^2}}
}\right].\qquad~~
\ea
The variables in this formal expression are constrained by the relations found above and which we summarize below:
\ba
\label{eqn:P_SNP2}
P(SNP)&\equiv& 
f_{00}\left( 
x_{0},\sigma_{0_D};\alpha_{ij}^{(k)},\beta\right)
=0.30\pm 0.001\\
\label{eqn:P_SNP|CHD2}
P(SNP\vert CHD)&\equiv& 
f_{02}\left( 
x_{0},\sigma_{0_D},\Delta_2^{thres};\alpha_{ij}^{(k)},\beta\right)
=0.20\pm 0.001
\\
P(t2DM\vert CHD)&\equiv& 
f_{22}\left( 
s_{2_D},\Delta_2^{thres};\alpha_{ij}^{(k)},\beta\right)
=0.09\pm 0.001~.
\label{eqn:P_t2DM|CHD2}
\ea
Here the subscript $D$ indicates properties of the data as derived form the model and constrained by these particular data, and the subscript $P$ indicates properties of the prior which are based on the knowledge inferred from data sets delivered by other experiments.
The results in Eqns.~(\ref{eqn:P_SNP2}) and (\ref{eqn:P_SNP|CHD2}) are of the same order as relative proportions found in other studies, respectively in Ref.~\cite{elahi_2008} and Refs.~\cite{dedoussis_2005,kammia_sxesh_3}. The result in Eqn.~(\ref{eqn:P_t2DM|CHD}) is a result of this study.

These functions depend on our knowledge of the rates in the model of the implicated pathways as well as on the statistical properties of the associated risk factor. 
However, from these three relations as constrained by the data, we can solve for three parameters only. Solving for the remaining parameters requires additional conditions for the statistical properties of the priors and biochemical parameters.
Nonetheless, the idea that the present study serves to introduce and which we here applied to one data set on one risk factor has been demonstrated, i.e. a) how to extract the statistical properties of the event from the corresponding phenomenological data and then b) from the statistical properties of the event how to extract biochemical information on the causal relations that link the event with the risk factor.

\section{Discussion}

In this manuscript we derive the probability of occurrence of CHD based on data in the presence of the SNP at the -308 position of the TNF-$\alpha$ gene. 
We first worked following a bottom-up approach.
Comparing different hypotheses for the statistical relation between the occurrences of SNP and CHD, we selected the working hypothesis on the basis of the Bayes factors. 
We showed that the data favour (although without strong evidence) 
the association of the SNP with the occurrence of CHD as well as 
the participation of the occurrence of t2DM in the causal relation. 
Using the  Bayes theorem, we computed the probability of the SNP conditional on the occurrence of CHD. 
We then worked following a top-down approach. We presented a schematic model for a simplistic description of the signalling pathway which relates the presence of the SNP with the emergence of CHD. The data contain information on equilibrium states of the several variables that describe the biochemical system and can thus be translated into a probability description.
We then computed the probability of CHD given that the SNP had occurred, using for the likelihood the probability previously computed. We expressed the result as a function of both the biochemical parameters of the model and the statistical parameters of the prior probability distributions. Other probabilities were also computed, which serve as constraints to the parameters.

In an upcoming study we will be exploring the idea further by integrating the sparse existing data on various population samplings. We will select the data for CHD given different risk factors, and for the SNP given different diseases. From the first selection we intend to extract the remaining statistical parameters, since the prior of CHD will be shared. 
Also a link should be established between this formalism and the CHD prediction estimates from a multivariable risk calculation \cite{wilson_1998}.
From the second selection we intend to extract the biochemical parameters of the 
signalling pathway. Although the prior of CHD will be shared, the biochemical system will grow in complexity and new rates will be introduced. We expect, however, that by exhausting the data sets available we will 
reach a balance of unknowns and equations that would allow us to solve the problem. Should this balance not be attained, we will resort to 
determining confidence levels for the unknown parameters based on simulations
\cite{girolami_2008, stumpf_2009,stumpf_2009b}. Whenever available, we will complement the study with temporal information to obtain reaction rates \cite{vourvouhaki_2007}. Ultimately we expect to be able to infer a universal law for gene mutation by systematising the various diseases into a comprehensive model of the signalling pathway.

\vspace{0.6cm}
\centerline{\bf {Acknowledgments}}
CSC is supported by Funda\c{c}\~ao para a Ci\^encia e a Tecnologia (FCT), SFRH/BPD/65993/2009.
The authors thank AM~Teixeira for a careful reading of the manuscript and PC~Aguiar, ME~Chollet and Z~Geitona for insightful comments. CSC also acknowledges the hospitality of the Astrophysical Sciences Department, Princeton University.
\hfill

\appendix
\section{Computation of the evidence}
\label{app:evidence}

In this Appendix we compute the evidence for the six hypotheses discussed in
Section~\ref{section:bottom-up}.
Hypothesis $H_{00}$ has only one free parameter, the probability that
the SNP occurred. This probability, describing a mutation process, is
assumed  to have a Poisson distribution characterized by the size of the
population $N$ and a mutation rate $\lambda.$ The probability of $n$
mutations is 
\ba
P(n\vert \lambda\wedge N)
=\exp[-\lambda N]
 {\left( \lambda N\right)^{n}\over n!}
\label{eqn:poisson}
\ea
with the mean number $\left<n\right>=\lambda N.$
For $n=N_{SNP}$ mutations in a sample of size $N$ and a uniform prior distribution for  $\lambda,$ $P\left(\lambda\right)=1,$ we find that
\ba
P\left( D_{SNP}\vert H_{00}\right)
&=&\int _{0}^{\infty} d\lambda~P\left( D_{SNP}\vert \lambda\wedge N\right)
 P\left(\lambda\vert H_{00}\right)\cr
&=&\int _{0}^{\infty} d\lambda
 ~\exp\left[-\lambda N\right]{\left(\lambda N\right)^{N_{SNP}}\over {N_{SNP}!}}
={1\over N}.
\label{eq:evidence_h00}
\ea

Hypothesis $H_{01}$ has two parameters, the probabilities that
the SNP occurred given the two values of the variable $CHD.$
There are two possible sources of SNP,
namely the population with CHD and the population without CHD. 
The presence of the SNP follows a binomial distribution where
$p_{01}$ is the frequency of the SNP for the case of
$CHD$ and $p_{0\bar 1}$ is the frequency of the SNP for the
case of $\overline{CHD.}$ 
It follows that 
\ba
P\left(D_{SNP}\vert H_{01}\right)
&=&\int dp_{01}\int dp_{0\bar 1}~
 P\left(D_{SNP}\vert p_{01}\wedge p_{0\bar 1}\wedge H_{01}\right)
  P\left(p_{01}\wedge p_{0\bar 1}\vert H_{01}\right)\cr
&=&\int dp_{01}~P\left(D_{SNP}\vert p_{01}\wedge H_{01}\right)
  P\left(p_{01}\vert H_{01}\right)\cr
 &&+\int dp_{0\bar 1}~P\left(D_{SNP}\vert p_{0\bar 1}\wedge H_{01}\right)
    P\left(p_{0\bar 1}\vert H_{01}\right).
\ea
Moreover,
assuming a uniform prior distribution probability for the
frequencies $p_{01}$ and $p_{0\bar 1}$ of the data on the SNP given
respectively the occurrence or non-occurrence of CHD 
\ba
P\left( p_{01}\vert H_{01}\right)=1, \quad
P\left( p_{0\bar 1}\vert H_{01}\right)=1,
\ea 
we find that
\ba
P\left(D_{SNP}\vert H_{01}\right)
&=&
\int dp_{01}~{N_{CHD}\choose N_{SNP,CHD}}
 p_{01}^{N_{SNP,CHD}}(1-p_{01})^{N_{\overline{SNP},CHD}}\cr
&&+\int dp_{0\bar 1}~{N_{\overline{CHD}}\choose N_{SNP,\overline{CHD}}}
 p_{0\bar 1}^{N_{SNP,\overline{CHD}}}(1-p_{0\bar 1})^{N_{\overline{SNP},\overline{CHD}}}\cr
&=&
{N_{CHD}\choose N_{SNP,CHD}}
{{N_{SNP,CHD}!N_{\overline{SNP},CHD}!}\over {\left( N_{CHD}+1\right)!}}\cr
&&+ {N_{\overline{CHD}}\choose N_{SNP,\overline{CHD}}}
  {{N_{SNP,\overline{CHD}}!N_{\overline{SNP},\overline{CHD}}!}
   \over {\left( N_{\overline{CHD}}+1\right)!}}
   \cr
&=&{1\over {N_{CHD}+1}}+{1\over {N_{\overline{CHD}}+1}}.
\ea

Similarly to $H_{01},$ hypothesis $H_{10}$ has two parameters, the
probabilities that the SNP occurred given the two values of the variable $t2DM.$
The evidence is given by the same
expression as that of hypothesis $H_{01}$ with the variable $CHD$
replaced by the variable $t2DM$ and under the analogous assumptions on
the corresponding frequency priors $p_{10}$ and $p_{\bar 1 0}.$

Hypothesis $H_{11}$ has four parameters, one for each state of the
variables $CHD$ and $t2DM.$ This hypothesis combines the two
hypotheses previously discussed which are assumed complementary,
thus being a case of a two-component 
hypothesis with probabilities
$\beta$ and $(1-\beta)$
\cite{fischer_2009}.
It follows that
\ba
P\left( D_{SNP}\vert H_{11}\right)
&=&\int dp_{01}\int dp_{10}~
\Bigl[
\beta P\left(D_{SNP}\vert p_{01}\wedge H_{11}\right)
  P\left(p_{01}\vert H_{11}\right)\cr
&&\quad+(1-\beta)P\left(D_{SNP}\vert p_{10}\wedge H_{11}\right)
  P\left(p_{10}\vert H_{11}\right)\Bigr]\cr
&&+
\int dp_{0\bar 1}\int dp_{\bar 1 0}~
\Bigl[
\tilde \beta P\left(D_{SNP}\vert p_{0\bar 1}\wedge H_{11}\right)
  P\left(p_{0\bar 1}\vert H_{11}\right)\cr
&&\qquad +(1-\tilde \beta)P\left(D_{SNP}\vert p_{\bar 10}\wedge H_{11}\right)
  P\left(p_{\bar 10}\vert H_{11}\right)\Bigr]
\ea
which yields
\ba
P\left( D_{SNP}\vert H_{11}\right)
&=&
\beta {N_{CHD}\choose N_{SNP,CHD}} 
{{N_{SNP,CHD}!N_{\overline{SNP},CHD}!}\over {\left(N_{CHD}+1\right)!}}\cr
&&\quad+(1-\beta){N_{t2DM}\choose N_{SNP,t2DM}} 
{{N_{SNP,t2DM}!N_{\overline{SNP},t2DM}!}\over {\left( N_{t2DM}+1\right)!}}
\cr
&&+
\tilde \beta {N_{\overline{CHD}}\choose N_{SNP,\overline{CHD}}} 
{{N_{SNP,\overline{CHD}}!N_{\overline{SNP},\overline{CHD}}!}
 \over {\left(N_{\overline{CHD}}+1\right)!}}\cr
&&\qquad+(1-\tilde \beta){N_{\overline{t2DM}}\choose N_{SNP,\overline{t2DM}}} 
{{N_{SNP,\overline{t2DM}}!N_{\overline{SNP},\overline{t2DM}}!}
 \over {\left( N_{\overline{t2DM}}+1\right)!}} 
\cr
&=&\beta {1\over {N_{CHD}+1}}
+(1-\beta){1\over {N_{t2DM}+1}}\cr
&&+
 \tilde\beta {1\over {N_{\overline{CHD}}+1}}
+(1-\tilde\beta){1\over {N_{\overline{t2DM}}+1}}.
\ea
Here $\beta=N_{CHD}/(N_{CHD}+N_{t2DM})$ is the probability that the
data were extracted from the pool of hypothesis $H_{01}$ 
and $\tilde \beta=N_{\overline{CHD}}
/(2N_{\overline{CHD},\overline{t2DM}}+N_{CHD,\overline{t2DM}}+N_{\overline{CHD},t2DM})$
the probability that the pool is that of the complement of $H_{01}.$
Similarly we define $(1-\beta)$ and $(1-\tilde \beta)$ from hypothesis
$H_{10}.$   

For Hypothesis $H_{11}^{chd}$ we have four parameters for combined states of the variables $CHD$ and $t2DM,$ namely
$CHD\wedge \overline{t2DM}$ and $\overline{CHD}\wedge \overline{t2DM}$ as the states which are conditional to non-occurrence of t2DM, and 
$t2DM\wedge CHD$ and $\overline{t2DM}\wedge CHD$ as the states which are conditioned to occurrence of CHD.  
The corresponding four frequencies are as follows:
$\tilde p_{01}$ is the frequency of SNP given
the occurrence of CHD and $\tilde p_{0\bar 1}$ the frequency of SNP given non-occurrence of CHD, both subject to non-occurrence of t2DM; 
$\tilde p_{11}$ is the frequency of SNP given that t2DM has
occurred and $\tilde p_{\bar 1 1}$ the frequency of SNP given that t2DM has not occurred, both subject to $CHD$ having occurred. For a uniform prior probability of these frequencies, we find that
\ba
P( D_{SNP}\vert H_{11}^{chd})
&=&\int d\tilde p_{01}\int d\tilde p_{11}~
\Bigl[
\gamma P(D_{SNP}\vert \tilde p_{01}\wedge H_{11}^{chd})
  P(\tilde p_{01}\vert  H_{11}^{chd})\cr
&&\quad+(1-\gamma)P(D_{SNP}\vert \tilde p_{11}\wedge H_{11}^{chd})
  P(\tilde p_{11}\vert H_{11}^{chd})\Bigr]\cr
&&+
\int d\tilde p_{0\bar 1}\int d \tilde p_{\bar 1 1}~
\Bigl[
\tilde \gamma P(D_{SNP}\vert \tilde p_{0\bar 1}\wedge H_{11}^{chd})
  P(\tilde p_{0\bar 1}\vert H_{11}^{chd})\cr
&&\qquad +(1-\tilde \gamma)
 P(D_{SNP}\vert \tilde p_{\bar 1 1}\wedge H_{11}^{chd})
  P(\tilde p_{\bar 1 1}\vert H_{11}^{chd})\Bigr] 
\ea
which yields
\ba
P( D_{SNP}\vert H_{11}^{chd})
&=&
\gamma {N_{CHD,\overline{t2DM}}\choose N_{SNP,CHD,\overline{t2DM}}} 
{{N_{SNP,CHD,\overline{t2DM}}!N_{\overline{SNP},CHD,\overline{t2DM}}!}
 \over {\left(N_{CHD,\overline{t2DM}}+1\right)!}}\cr
&&\quad+(1-\gamma){N_{CHD,t2DM}\choose N_{SNP,CHD,t2DM}} 
{{N_{SNP,CHD,t2DM}!N_{\overline{SNP},CHD,t2DM}!}
 \over {\left( N_{CHD,t2DM}+1\right)!}}
\cr
&&+
\tilde \gamma 
{N_{\overline{CHD},\overline{t2DM}}\choose N_{SNP,\overline{CHD},\overline{t2DM}}} 
{{N_{SNP,\overline{CHD},\overline{t2DM}}!N_{\overline{SNP},\overline{CHD},\overline{t2DM}}!}
 \over {\left(N_{\overline{CHD},\overline{t2DM}}+1\right)!}}\cr
&&\qquad+(1-\tilde \gamma)
{N_{CHD,\overline{t2DM}}\choose N_{SNP,CHD,\overline{t2DM}} } 
{{N_{SNP,CHD,\overline{t2DM}}!N_{\overline{SNP},CHD,\overline{t2DM}}!}
\over {\left( N_{CHD,\overline{t2DM}}+1\right)!}} 
\cr
&=& \gamma{1\over {N_{CHD,\overline{t2DM}}+1}}
+(1-\gamma){1\over {N_{CHD,t2DM}+1}}\cr
&&+ \tilde\gamma{1\over {N_{\overline{CHD},\overline{t2DM}}+1}}
+(1-\tilde\gamma){1\over {N_{CHD,\overline{t2DM}}+1}}.
\label{eqn:h_11^chd}
\ea
Here $\gamma=N_{CHD,\overline{t2DM}}/N_{CHD}$ and 
$\tilde \gamma=N_{\overline{CHD},\overline{t2DM}}
/N_{\overline{t2DM}}.$

Similarly to $H_{11}^{(chd)},$ hypothesis $H_{11}^{(t2dm)}$ has four
parameters for combined states of the variables $CHD$ and $t2DM,$ namely
$t2DM\wedge \overline{CHD}$ and $\overline{t2DM} \wedge \overline{CHD}$ as the states conditional on non-occurrence of CHD, and $CHD\wedge t2DM$ and $\overline{CHD}\wedge t2DM$ as the states conditional on occurrence of t2DM.
The frequencies are analogously defined to those of hypothesis $H_{11}^{(chd)}.$
The evidence is given by the same expression as that of
hypothesis $H_{11}^{(chd)}$ with 
$\tilde p_{01}$ and $\tilde p_{0\bar1}$ replaced by 
the frequency of SNP, subject to non-occurrence of CHD, given the occurrence or non-occurrence of t2DM respectively 
$\tilde p_{10}$ and $\tilde p_{\bar 1 0},$ 
and $\tilde p_{\bar 1 1}$ replaced by $\tilde p_{1\bar 1}.$ Analogously
$\gamma$ is replaced by 
$N_{\overline{CHD},t2DM}/N_{t2DM}$ and 
$\tilde \gamma$ by 
$N_{\overline{CHD},\overline{t2DM}}
/N_{\overline{CHD}}.$

\section{Calculation of $P(SNP\vert H_{00})$}
\label{h00}

In this Appendix we compute for the purpose of comparison 
the probability of occurrence of SNP for hypothesis $H_{00}.$
Starting from Eqn.~(\ref{eqn:poisson}) 
and using the Bayes theorem, we find for the posterior probability of
$\lambda$ that
\ba
P\left(\lambda\vert D_{SNP}\wedge N\right)
={{P\left( D_{SNP}\vert \lambda \wedge N\right)P\left(\lambda\vert N\right)}
  \over {P\left(D_{SNP}\vert N\right)}}.
\ea
The normalizing constant is the evidence computed in Eqn.~(\ref{eq:evidence_h00}).
Given the data and for $P(SNP\vert \lambda \wedge H_{00})=\lambda,$
the probability of a mutation in a population of size $N$ is 
\ba
P\left(SNP\vert H_{00}\right)
&=&\int d\lambda~P\left(SNP\vert \lambda \wedge H_{00}\right)
  P\left(\lambda\vert D_{SNP} \wedge H_{00}\right)\cr
&=&\int d\lambda~\lambda
  N\exp\left[-\lambda N\right]{\left(\lambda N\right)^{N_{SNP}}\over {N_{SNP}!}}
={{N_{SNP}+1}\over {N}}
\ea
which for $N_{SNP}=184$ 
and $N=683$ 
yields $P\left(SNP\wedge H_{00}\right)=0.27.$ Despite the small difference between the values derived from the two hypotheses (which might be considered insignificant given the observational errors which are of order $1/\sqrt{N}\approx 0.038$), the fact that there is a difference highlights the relevance of hypothesis testing before committing to a probability which will act as likelihood in subsequent calculations.

\end{document}